\documentstyle[twocolumn,prl,aps]{revtex}

\begin{document}
\draft
\twocolumn[
\hsize\textwidth\columnwidth\hsize\csname @twocolumnfalse\endcsname

\draft
\title{
	Potential energy landscape and long time dynamics
	in a simple model glass
      }
\author{
	L.~Angelani$^{1}$,
	G.~Parisi$^{2}$,
	G.~Ruocco$^{1}$ and 
	G.~Viliani$^{3}$,
       }
\address{
	 $^{1}$ 
	 Universit\'a di L'Aquila and Istituto Nazionale di Fisica 
	 della Materia, I-67100, L'Aquila, Italy. \\
	 $^{2}$ 
	 Universit\'a di Roma {\rm La Sapienza} and Istituto Nazionale di 
	 Fisica Nucleare, I-00185, Roma, Italy. \\
	 $^{3}$ 
	 Universit\'a di Trento and Istituto Nazionale di Fisica
	 della Materia, I-38050, Povo, Trento, Italy.\\
	}
\date{\today}
\maketitle
\begin{abstract}
We analyze the properties of a Lennard-Jones system at the level of the 
potential energy landscape.
After an exhaustive investigation of the topological features of the landscape 
of the systems, obtained studying small size sample, we describe the dynamics
of the systems in the multi-dimensional configurational space by a simple
model. This consider the configurational space as a connected network of 
minima where the dynamics proceeds by jumps described by an appropriate 
master equation. Using this model we are able to reproduce the long time 
dynamics and the low temperature regime. We investigate both the equilibrium 
regime and the off-equilibrium one, finding those typical glassy behavior
usually observed in the experiments such as: {\it i)} stretched exponential
relaxation,  {\it ii)} temperature-dependent stretching parameter,  {\it iii)}
breakdown of the Stokes-Einstein relation, and  {\it iv)} appearance of a 
critical temperature below which one observes deviation from the 
fluctuation-dissipation relation as consequence of the lack of equilibrium
in the system.

\end{abstract}
]

\section{Introduction}

The landscape paradigm \cite{landscape} is a very useful point of view for
the study of glassy systems. The detailed analysis of the free energy or
potential energy surfaces allows us to get insight into the rich
phenomenology exhibited by glass-forming liquids in the supercooled phase,
around the glass transition region and in the low temperature glassy regime. 
While the investigation of the free energy surface is a very hard task 
starting from the microscopic description of the system, since the landscape 
details are very strongly temperature dependent, the description of the 
potential energy landscape is a much more tractable problem, and is a good 
starting point to investigate properties at not-too-high temperatures. 
The trajectory of the representative point of the system in the 
configurational phase space can be viewed as a path in the multidimensional 
potential energy surface. The dynamics is strongly influenced by the 
topography of the landscape: local minima, barriers height, attraction 
basins and further topological features. In recent years it has been show 
that the details of the potential energy surface are of great importance 
in determining the properties of many systems exhibiting glassy behavior, 
like glass-forming liquids, protein folding, atomic cluster or evolutionary 
biological models.

In this paper we numerically investigate the low temperature dynamical 
properties of a simple system, i.e. a monoatomic Lennard-Jones system, 
through an analysis of its multidimensional potential energy surface
and a simple model for the loe temperature dynamics. In the first part 
(Section II), by studying small sample size, we give a comprehensive 
description of the potential energy landscape of the systems: minima, 
barriers, reaction paths, saddle points, and determine statistical 
distributions and cross-correlations among the analyzed quantities. 
In the second part (Section III) we use this information to set up a 
simple model for the study of the long-time relaxation dynamics of the 
system; the model consists of a connected network of potential energy 
minima with a jump dynamics described by an appropriate master equation. 
This allows us to get information on the behavior of the system for times 
long enough that a direct Molecular Dynamics (MD) simulation is not 
feasible \cite{lettera}. After a static (thermodynamical) test of the model, 
we determine the dynamical equilibrium properties and also the off-equilibrium
ones, and discuss the results. In the last Section we report the conclusions.


\section{Potential energy landscape}

We numerically investigate the topology of the potential energy hypersurface
of a Lennard-Jones $6-12$ system of $N$ interacting particles in a cubic box
with periodic boundary conditions. The pair potential is: 
\begin{equation}
V_{LJ}(r)=4\epsilon \ \left[ \left( \frac \sigma r\right) ^{12}-\left( \frac 
\sigma r\right) ^6\right] \ ,
\end{equation}
with $r$ the Euclidean distance between two particles. The physical
parameters $\sigma $ and $\epsilon $ are choose to describe an Argon system: 
$\sigma =0.3405$ nm and $\epsilon /K_B=125.2$ K ($K_B$ is the Boltzmann
constant). In order to have an exhaustive description of the energy
landscape, we investigate small size systems, with a number of particles 
$N<30$. As we shall see later, such small systems can nevertheless exhibit
complex enough behavior.

Due to the small size of the system, the range of interactions among the
particles is of the same order as the box length. It is then much more
appropriate to use a multi-image method instead of the usual minimum image
method \cite{ALLEN}, in which each particle interacts only with the nearest
image (generated by the periodic boundary conditions) of all other
particles. In our case many images contribute, and it is necessary to
consider them in the calculation of the interactions. The choice we made
introduces an angular dependence in the pair potential which, however, is
negligible with respect to the radial one. The simulated density is $\rho
=4.2\cdot 10^{-2}$ mol/cm$^3$, which was obtained by Demichelis et al. 
\cite{DEMICHE} as the smallest value at which a Lennard-Jones system of 
$N=864$ Argon particles is stable in the glassy state after a rapid quench 
from high temperature. 

\subsection{Minima}

As a first step we search for the potential energy minima. We used a
modified conjugate gradient method starting from high temperature
configurations obtained during a MD simulation. In this way we find the so
called ``inherent configurations'', corresponding to local minima. New
minima are identified by their potential energy values. We analyzed systems
with $N=11\div 29$ and, for each $N$, we stopped the search when the rate at
which new low energy minima are found 
(for example in lowest first third of the full energy range) 
is smaller than a given number (about $10^{-4}$). 
In this way we were able to obtain a good classification
of the low energy minima. The number of found minima, ${\cal N}$, does not
show a clear and well defined dependence on $N$, contrary to the case of
small clusters \cite{DALDOSS}, so that it is not possible to give an
estimate of the coefficient $\alpha $ in the expected exponential growth: 
${\cal N}\propto \exp (\alpha N)$ (for clusters $\alpha \sim 1$). We observe
indeed for some values of $N$ a strong tendency of the system to fall always
in the same minima. In this case an exhaustive research of the inherent
structures is a very hard numerical task. this explains the not clear
dependence found.

Once most minima are classified, we analyze their features: energy, static
structure factor, curvature and distances.

\subsubsection{Energy}

The first important information about an inherent configuration is its
potential energy value. From the energy distribution it is sometimes
possible to recognize the crystalline-like configurations. Usually one finds
an evident gap between the lowest energy minima, the crystalline-like ones,
and the other ones with higher energies, as can be seen in Fig. $1$ where we
report the distribution of minima energies per particle for $N=29$. In other
cases the situation is not so clear and it is useful to use a different
method to characterize the nature of an inherent configuration; the most
useful quantity in determining the spatial order structure of a minimum is
its static structure factor, which allows (imperfect) crystal-like
configuration also of high energy to be identified.

\subsubsection{Static structure factor}

In order to classify the spatial distribution of particle in a given 
minimum of the potential energy we use the static structure factor: 
\begin{equation}
{\hat S}({\vec q})=\frac 1N|\sum_j\exp (i{\vec q}\cdot {\vec r}_j)|^2\ .
\end{equation}
Due to the finite size of the system the allowed ${\vec q}$ vectors are of
the form ${\vec q}=(2\pi /L)\ {\vec n}$, with ${\vec n}=(n_x,n_y,n_z)$ an
integer vector. We define the quantity 
\begin{equation}
S(q)=\frac 1{n_q}\sum_{\vec q \in \{q,\Delta q\}}{\hat S}({\vec q})\ ,
\end{equation}
where $\sum_{\vec q \in \{q,\Delta q\}}$ is a sum over the $n_q$ 
vectors with modulus within $q \pm \Delta q$.
For a pure crystalline configuration of $N$ particles $S(q)$
consists of Bragg peaks, and its value at the peaks is $S_{max}=N$. For
amorphous configurations $S(q)$ does not present a well defined peak
structure and the highest value is $S_{max}\sim 2\div 3$, obtained for a $q$
value of the order of the inverse mean distance of two near particles. For
small sized systems, like those analyzed here, there are intermediate
situations and we use the criterion $S_{max}<N/2$ to determine the amorphous
nature of an inherent structure.

\subsubsection{Curvature}

Another important property of a minimum is its overall curvature $c$, 
defined as the determinant of the Hessian  $\Phi^{^{\prime \prime }}$
of the potential energy function $\Phi$: 
\begin{equation}
c=\det (\Phi^{^{\prime \prime }})\ ,
\end{equation}
The eigenvalues of the Hessian matrix are proportional to the squared
vibrational eigenfrequencies.
In the inset of Fig. 1 we show the distribution of 
\begin{equation}
\gamma=\frac 1{3N-3}\ {\rm Log}_{10}\ (c/m)^{1/2}\ ,
\end{equation}
where $m$ is the mass of the particles 
(we use the Argon's value $m=40$ amu).
The quantity $\gamma$ is thus 
proportional to the sum of the logarithms of the frequencies of
normal vibrational modes 
\begin{equation}
\gamma=\frac 1{3N-3}\ \sum_\alpha \ \ln \ \omega _\alpha\ ,
\end{equation}
with $\alpha =1,...,N-3$ (the three zero frequencies corresponding to rigid
translations have been eliminated from the sum). As can be seen in Fig. 1, 
the highest $\gamma$ values correspond to the minima with lowest energy, 
i.e. the crystalline-like minima which are narrower and deeper than
the other packing structures (see also Sec. II-C).

\subsubsection{Stress tensor}

For each minimum we determined the off-diagonal part of the microscopic
stress tensor: 
\begin{equation}
\left\{ 
\begin{array}{rcl}
\displaystyle \sigma ^{zx} & = & \displaystyle -\sum_{i<j}V_{LJ}^{^{\prime
}}(r_{ij})\frac{z_{ij}x_{ij}}{r_{ij}}\ , \\ 
&  &  \\ 
\displaystyle \sigma ^{xy} & = & \displaystyle -\sum_{i<j}V_{LJ}^{^{\prime
}}(r_{ij})\frac{x_{ij}y_{ij}}{r_{ij}}\ , \\ 
&  &  \\ 
\displaystyle \sigma ^{yz} & = & \displaystyle -\sum_{i<j}V_{LJ}^{^{\prime
}}(r_{ij})\frac{y_{ij}z_{ij}}{r_{ij}}\ ,
\end{array}
\right.  \label{tsfo1}
\end{equation}
where $x_{ij},y_{ij},z_{ij}$ are the components of ${\vec r}_{ij}=
{\vec r}_i-{\vec r}_j$; 
these quantities will be useful in determining the shear
viscosity. The form of the stress tensor we use is the $q=0$ extrapolation
of the $q$-dependent expression \cite{BALUCANI} 
\begin{eqnarray}
\sigma _{\alpha ,\beta }({\vec q}) &=&\sum_i\left[ mv_{i,\alpha }v_{i,\beta
}-\sum_{i<j}\frac{r_{ij,\alpha }r_{ij,\beta }}{r_{ij}^2}P_q(r_{ij})\right]
\cdot  \nonumber  \label{tsfo2} \\
&&\cdot \exp (i{\vec q}\cdot {\vec r}_i)\ ,
\end{eqnarray}
with 
\begin{equation}
P_q(r)=rV_{LJ}^{^{\prime }}(r)\frac{1-\exp (i{\vec q}\cdot {\vec r})}{i{\vec 
q}\cdot {\vec r}}\ .
\end{equation}
The index $i$ and $j$ refer to particles, while $\alpha $ and $\beta $ label
the spatial axes $x,y,z$. In (\ref{tsfo1}) we have omitted the kinetic
term, not well defined in an inherent configuration; our hypothesis is that
also this truncated form is able to well describe the relaxation processes.

\subsubsection{Distances}

We now turn to the relationships among the minima; in particular we determined
the mutual distances in the $3N$-dimensional configuration space; in this
regard it is important to take into account all the symmetry operations of
the problem. Indicating with ${\underline{r}}_a=({\vec r}_1^a,...,
{\vec r}_N^a)$ the $3N$ coordinates of the particles in the minimum $a$, 
we define the distances $d_{ab}$ between minima $a$ e $b$: 
\begin{equation}
d_{ab}=\min_{{\small {T,R,\pi }}}(|{\underline{r}}_a-{\underline{r}}_b|)\ ,
\label{dista}
\end{equation}
where the minimization is made with respect to the continuous translations 
($T$), discrete rotations and reflections ($R$), and permutations ($\pi $) of
the particles.

The minimization over the continuous translations ($T$) is done by putting,
sequentially, each particle of $a$ in the same place as each one of $b$. 
The minimization over the rotations and
reflections ($R$) is carried out by considering the $48$ symmetry operations
of the cubic group; the minimization over particle permutation ($\pi $) is
apparently an hard task to solve, as one should consider all the $N!$
possible configurations in a direct calculation, but this is not the case
actually. The problem is of polynomial type, i.e. the time needed to
find the optimum solution grows as a polynomial function of the size, $N$,
of the system (on the other hand non-polynomial problems require an
exponential computational time, i.e. the travel salesman problem). The
optimization problem we need to solve is a bipartite matching problem, which
can be done in very short computational time by using an appropriate
algorithm.

\subsection{Barriers}

A very important topological quantity in determining the dynamical behavior
of the system is the energy barrier experienced by the system in traveling
from one minimum to another. At first sight it might seem that it is
accurate enough to evaluate the barrier along the straight path joining two
minima in the $3N$-dimensional space. However, as evidenced by Demichelis et
al. \cite{DEMICHE}, this produces in most cases much higher barriers than an
evaluation along the least action path, indicating that the straight path
approximation is often not good. We have then determined the least
action path for each pair of minima $a$ and $b$, which is defined as the
path that minimizes the action functional 
\begin{equation}
S[\ell ]=\int_\ell ds\sqrt{2m[\Phi ({\underline{r}}(s))-\Phi _0]}\ ,
\end{equation}
where $\ell $ is a generic path between the minima, $s$ is the curvilinear
coordinate and $\Phi _0=\min \{\Phi _a,\Phi _b\}$. This functional problem
is simplified by dividing the path into a finite number of intervals 
(typically we use $n=16$ intervals) and by minimizing the action function 
with respect to the extreme of the $n$ segments constrained to move in 
hyperplanes perpendicular to the straight path. The highest potential-energy 
value along the least action path determines the barrier height and identifies 
the saddle point.

We have only analyzed the system with $N\leq 17$, due to the very long
computational times needed in the cases with $N>17$ where there are too many
pairs of minima to take into account. We report the results for the largest
system analyzed, $N=17$ with ${\cal N}=38$ minima. In Fig. $2$ we show an
example of the potential energy profile along the straight path (dashed
line) and along the least action path (full line) between two minima. It is
evident that the energy barrier is significantly lower in the latter case,
although the two paths are not very distant in configuration space. Similar
results are obtained for the other analyzed paths. Sometimes it happens that
two minima are not directly connected, in the sense that the least action
path joining them crosses a third minimum, and a non trivial connectivity
among the minima emerges (in the analyzed system we find that each minimum 
on an average is directly connected to $20$ other minima).
For each saddle point along the least action path we determine the main
properties: energy, curvature, down ''frequency''. In Fig. $3$ is reported
the energy distribution of the barriers $\Delta\Phi_{bar.}$. 
The curvature is defined as the
absolute value of the determinant of the Hessian of the potential energy
evaluated at the saddle: 
\begin{equation}
c_{sad.}=|\det \Phi _{sad.}^{^{\prime \prime }}|\ ,
\end{equation}
In the inset of Fig. $3$ we show the distribution of the quantity 
\begin{equation}
\gamma _{sad.}=\frac 1{3N-3}\ {\rm Log}_{10}\ (c_{sad.}/m)^{1/2}\ .
\end{equation}
The down ``frequency'' ${\tilde \omega }_{sad.}$ is 
defined as the square root of the absolute value of the
down curvature along the least action path: 
\begin{equation}
{\tilde \omega }_{sad.}^2=-\frac{{\underline{v}}\ 
\Phi _{sad.}^{^{\prime \prime
}}\ {\underline{v}}}{|{\underline{v}}|^2}\ ,
\end{equation}
where ${\underline{v}}$ is the tangent vector to the least action path at
the saddle point. 

\subsection{Correlations}

In order to have a full statistical description of the potential energy
landscape, it is useful to investigate, besides the distributions of the
different quantities, also their cross-correlations. We have then determined
the linear correlation coefficient 
\begin{equation}
r(x,y)=\frac{\sum_i(x_i-{\overline{x}})(y_i-{\overline{y}})}{[\sum_i(x_i-
{\overline{x}})^2\ \sum_j(y_j-{\overline{y}})^2]^{1/2}}\ 
\end{equation}
for all the measured quantities $x$ and $y$. In Table I are reported the
values obtained, together with the $\log -\log $ correlations, in order to
evidence possible power laws. In the first column we report the number of
particles of the system analyzed (only for $N=17$ we have determined the
least action paths and all the related quantities). It emerges that,
substantially, energy difference and distance among the minima are not
correlated, indicating that the topological structure of the inherent
configurations is not energy-correlated. A weak correlation is observed
between energy and curvature at stationary points (minima and saddle
points). In Fig. $4$ we show the cross-correlation between the energies of
the minima and their curvatures. 
An interesting correlation is observed between barrier energies 
$\Delta\Phi_{bar.}$ and distances among minima $d_{ab}$ (Fig. $5$), 
with a nearly linear correlation in double $\log $ scale 
(line in the figure).

We conclude the analysis of the energy landscape by determining the entropy
ratio $R$, defined as the ratio between the curvature of saddle points and
that of the related minima 
\begin{equation}
R=\frac{|\det \Phi _{sad.}^{^{\prime \prime }}|}{\det \Phi
_{min.}^{^{\prime \prime }}}\ .
\end{equation}
This quantity gives a quantitative measure of the ability of the system in
finding the right path to reach another minimum. If $R\sim 1$ there are no
entropic hindrances, while if $R\gg 1$ these effects become relevant, as the
narrowness (higher value of the curvature) at the saddle makes the least
action path toward that specific minimum unfavorable with respect to other 
escape route. For all the minimum-saddle-minimum triplets we have evaluated 
$R$; the majority of the values is in the range $10^{-2}\div 10$, in 
qualitative agreement with the results found in Lennard-Jones clusters 
\cite{DALDOSS}.


\section{Model for the dynamics}

The investigation of the properties of glass-former liquids at the level of
the energy landscape, allows us to introduce some approximations in the
dynamics of the system. We define a simplified model which is able to
capture the long time behavior of the system, and which consists of a
connected network of potential energy minima with a jump dynamics among them
described by an appropriate master equation \cite{lettera}.

The basic idea is quite simple. A glass is represented by a configurational
point confined in a very small region of the accessible phase space and in
the zero temperature limit (neglecting quantum effects) all the atoms are
frozen in well defined positions, corresponding to some mechanically
metastable state. When the temperature is raised jumps among different
mechanically stable positions become possible. At finite and not-too-high
temperature we assume the dynamically relevant processes are the following:
a short time dynamics dominates by small vibrations around stable positions
(this dynamics can be described within the harmonic approximation by
diagonalizing the dynamical matrix), and a long time dynamics consisting of
collective (involving many atoms) jumps among different stable positions.
The main hypothesis we make is that there is a substantially clear
separation of time scales between the two dynamical processes. This
characterization of the dynamics is a good approximation at not-too-high
temperature. By increasing the temperature, anharmonic effects become
relevant to the vibration around the local minimum and, moreover, a clear 
time scale separation between fast vibrational and slow jumps dynamics is 
no longer possible. In a recent work \cite{SCHRO} the validity of this 
hypothesis has been verified in a Lennard-Jones binary mixture, giving 
a very good agreement with a direct MD investigation. To sum
up, our model, which is expected to capture the physics of the system at low
temperature, is based on two main hypotheses:

\begin{enumerate}
\item  clear cut difference between vibrational dynamics at short time and
dynamics of collective jumps at long time;

\item  description of the long time dynamics through a master equation, with
the transition rates that depend on the topological properties of the
potential surface.
\end{enumerate}

The main advantages of the introduced model with respect to the usual MD
computations are:

\begin{itemize}
\item  we can avoid in a simple way the crystallization process, that always
takes place in one component LJ systems, as we
do not consider the crystalline minima in setting up the network.

\item  we can study in a direct way the low temperature properties, where
usually the very long relaxation times require very long computational time.
In MD the computational times are proportional to the physical times, while
in the model introduced here the computational times are those needed to find 
the eigenvalues and eigenvectors of the transition matrix, independent of
temperature;

\item  it is possible to evidence the relationships between the energy
landscape and the behavior of the system.
\end{itemize}

To be more specific, the model is a connected network of potential energy
minima and the master equation governing the jumps dynamics is: 
\begin{equation}
\frac{dp_a}{dt}(t;b,t_0)=\sum_c\ W_{ac}\ p_c(t;b,t_0)\ ,  \label{me}
\end{equation}
where $p_a(t;b,t_0)$ is the probability that the system is at minimum $a$ at
time $t$, if it was at minimum $b$ at time $t_0$. The off-diagonal elements
of the matrix $W$ are the transition rates. The diagonal elements are fixed
by the condition 
\begin{equation}
\sum_aW_{ac}=0\ .
\end{equation}
In order to obtain an asymptotic behavior that reproduces the right
Boltzmann weight the occupation probability must satisfy:
\begin{equation}
\lim_{t\rightarrow \infty }p_a(t;b,t_0)=p_a^0\equiv \frac 1{{\cal Z}}(\det
\Phi _a^{^{\prime \prime }})^{-1/2}\ \exp (-\beta \Phi _a)\ ,
\end{equation}
(${\cal Z}$ is such that $\sum_ap_a^0=1$ and the pre-exponential factor
follows from the harmonic vibration in each minimum), and the transition 
matrix  $W$ must satisfy the detailed balance relation: 
\begin{equation}
W_{ab}\ p_b^0=W_{ba}\ p_a^0\ .
\end{equation}
The solution of the master equation is easily expressed in terms of
eigenvalues $\lambda _n$ and eigenvectors $\alpha _a^{(n)}$ ($n=1,...,M$,
with $M$ matrix dimension) of $W$: 
\begin{equation}
p_a(t;b,t_0)=(p_b^0)^{-1}\ \sum_n\ \alpha _a^{(n)}\ \alpha _b^{(n)}\ \exp
[\lambda _n(t-t_0)]\ .  \label{me2}
\end{equation}
In the numerical calculus it is more convenient to express the solutions in
terms$_{}$ of the eigenvectors of a new symmetric matrix $w_{ab}=W_{ab}\
(p_b^0/p_a^0)^{1/2}$ (whose eigenvalues coincide with those of $W$): 
\begin{equation}
p_a(t;b,t_0)=(p_a^0/p_b^0)^{1/2}\ \sum_n\ e_a^{(n)}\ e_b^{(n)}\ \exp
[\lambda _n(t-t_0)]\ .  \label{me3}
\end{equation}
The model is well defined once we give an appropriate form to the transition
matrix $W$. In order to determine the transition rates let us analyze the
problem of escape from a metastable state; a useful point of view for
systems with many degree of freedom is the description in term of a few
relevant coordinates. This reduction is possible whenever there are few
reaction coordinates with characteristic evolutionary times longer than
those of the other degrees of freedom, which act as effective terms on the
relevant coordinates, i.e. like noise and viscous terms. We suppose this is
the case of our system whenever the temperature is not too high (the
analysis of reaction paths made by Demichelis et al. \cite{DEMICHE} supports
this hypothesis). Handling the problem as a Markovian-Brownian 
$d$-dimensional motion in the overdamped friction regime we obtain the form 
\cite{RISKEN} 
\begin{equation}
W_{ab}=\frac{\tilde \omega _{sad.}^2}\mu \left[ \frac{\det \Phi _b^{^{\prime
\prime }}}{|\det \Phi _{sad.}^{^{\prime \prime }}|}\right] ^{1/2}\ \exp \left[
-\frac{\Phi _{sad.}-\Phi _b}{K_BT}\right] \ ,  \label{wab}
\end{equation}
where $\tilde \omega _{sad.}$ is the down ``frequency'' at the saddle point
and $\mu $ is a friction constant that determines the time scale (its value
is fixed by a comparison with MD in the allowed temperature region). 

All the characteristics of the model (properties of the connected network
and parameters in the transition rates) are inferred from the computed
properties of the potential energy landscape. We use the values of the $N=29 
$ system to determine local minima properties (energy, curvature, stress
tensor) and those of $N=17$ system to determine connectivity properties
(energy and curvature of saddle points, distances and connectivity among the
minima). The values are extracted from the distribution found in the
previous section in the following way:

\begin{enumerate}
\item  we extract $M$ energy values of the minima from the distribution of 
$N=29$ system (we exclude the crystalline-like configurations);

\item  we assign to each minimum a value of curvature $c_a=\det \Phi
_a^{^{\prime \prime }}$ extracted from a bivariate distribution, thanks to
the cross-correlation between energy and curvature; a stress tensor value is
also extracted for each minimum;

\item  for each minimum we randomly (in the analysis of the energy landscape
we have found no correlation between energies and distances among minima)
extract $20$ minima connected to it, as obtained on an average for 
the system $N=17$;

\item  we define a connection matrix $\kappa _{ab}$, 
containing the minimum steps, i.e. the number of minima crossed,
necessary to go from $a$ to $b$;
the distance matrix $d_{ab}$ is $\kappa _{ab}$ times the value 
extracted from the distribution of the distances among connected 
minima for $N=17$;

\item  for each pair of directly connected minima we determine the energy
barriers $\Delta\Phi_{bar.}$ from the value of the distance $d_{ab}$: 
$\Delta\Phi_{bar.}=A\ d^\alpha $ ($A\simeq 10^5$ and $\alpha \simeq 3.7$, as
determined for $N=17$ system, Fig. 5);

\item  we assign a curvature value 
$c_{sad.}=|\det \Phi _{sad.}^{^{\prime \prime }}|$ 
and a down ``frequency'' $\tilde \omega _{sad.}$ to each saddle point,
from bivariate distribution.
\end{enumerate}

We obtain in this way a set of parameters that describe the model. In order
to have a good statistic description we considered different extractions of
the parameters and the measured quantities were obtained by averaging over
the extractions.

\subsection{Test}

Before studying the dynamical properties of the model, we concentrate on the
static behavior obtained as asymptotic solution of the master equation. In
this static regime we can determine, in a statistical mechanical approach,
the configurational partition function 
\begin{equation}
{\cal Z}(\beta )=\int d^{3N}r\ \exp [-\beta \Phi ({\vec r}_1,...,{\vec r}_N)]
\ ,
\end{equation}
By using the approximation based on the hypothesis of short time local
harmonic vibrations around a minimum, and long time collective jumps among
different minima, we obtain 
\begin{equation}
{\cal Z}(\beta )\sim \sum_a{\cal Z}_a^{(harm.)}(\beta )\ \exp [-\beta \Phi
_a]\ ,
\end{equation}
where $a$ labels the minima and ${\cal Z}_a^{(harm.)}$ is the contribution
of harmonic vibrations around minimum $a$. This form of the configurational
partition function emerges in the model as the exact infinite-time limit.
The harmonic term is easy to calculate, being a $3N$-dimensional Gaussian
integral: 
\begin{eqnarray}
{\cal Z}_a^{(harm.)}(\beta ) &=&\int d{\underline{r}}\exp \left[- \frac \beta 
2\ {\underline{r}}\ \Phi _a^{^{\prime \prime }}\ {\underline{r}}\right] 
\nonumber \\
\ &=&(2\pi)^{3N/2}\ \beta ^{-3N/2}\ (\det \Phi _a^{^{\prime \prime
}})^{-1/2}\ ,
\end{eqnarray}
where ${\underline{r}}=(r_1,...,r_{3N})$, and ${\underline{r}}\ \Phi
_a^{^{\prime \prime }}\ {\underline{r}}=\sum_{l,m}r_l\ (\Phi _a^{^{\prime
\prime }})_{lm}\ r_m$. 
We then obtain the approximated partition function as: 
\begin{equation}
{\cal Z}(\beta )\sim c\ \beta ^{-3N/2}\ \sum_a\ (\det \Phi _a^{^{\prime
\prime }})^{-1/2}\ \exp (-\beta \Phi _a)\ ,  \label{zapp}
\end{equation}
from which the thermodynamical quantities can be derived, for example for
the energy: $E(\beta )=- \partial _\beta \ $ln\ ${\cal Z}$. 
To check the reliability of the model we compare the quantities calculated
from (\ref{zapp}) with those obtained through MD computation. In Fig. $6$ we
show the potential energy as obtained from the model (lines) and from MD
(circles). The MD data are obtained in the following way. Starting from high
temperature we quench rapidly the system to low temperature, entering in a
glassy state; we then increase the temperature up to liquid phase (open
circles). The system is subsequently slowly cooled, entering in the
supercooled regime ($100-70$ K) and obtaining at the end the crystal through
a first order transition (full circles). The lines represent the energies
determined from the model by taking into account all the minima (dotted
line) and only the glassy ones (full line). A good quantitative agreement is
obtained between MD and the model as far as the temperature is lower than
about $150$ K, a temperature in the liquid phase well above the melting
point ($T_m\sim 80$ K). This result supports the correctness of the
approximation of local-vibration / collective-jumps in the description of a
glass-former at not-too-high temperature. This static test is a good
starting point to extend the analysis to the dynamical regime.

\hfill

\subsection{Equilibrium properties}

We now determine the dynamical equilibrium properties of the model. We
denote with ${\cal O}({\underline{r}}(t),{\underline{r}}(0))$ a generic
observable which depends on collective coordinates ${\underline{r}}$ at time 
$t$ and at initial time $t=0$. We define the statistical average value of 
${\cal O}$ in the model as 
\begin{equation}
<{\cal O}(t)>=\sum_bp_b^0\ \sum_a{\cal O}_{ab}\ p_a(t;b,0)\ ,  \label{oss1}
\end{equation}
where ${\cal O}_{ab}$ is the value of ${\cal O}$ evaluated at the minimum
configurations $a$ and $b$: ${\cal O}_{ab}={\cal O}({\underline{r}}_a,
{\underline{r}}_b)$. In terms of the eigenvalues and eigenvectors of the
transition matrix $W$ we have: 
\begin{equation}
<{\cal O}(t)>=\sum_n\exp (\lambda _nt)\ \sum_{a,b}{\cal O}_{ab}\ \alpha
_a^{(n)}\ \alpha _b^{(n)}\ ,  \label{oss2}
\end{equation}
or in terms of the eigenvectors of the symmetric matrix $w$: 
\begin{equation}
<{\cal O}(t)>=\sum_n\exp (\lambda _nt)\ \sum_{a,b}{\cal O}_{ab}\ (p_a^0\
p_b^0)^{1/2}\ e_a^{(n)}\ e_b^{(n)}\ .  \label{oss3}
\end{equation}
In the following we report a detailed analysis of the equilibrium dynamics
for a network of $400$ minima, averaging over $50$ different extractions of
the parameters that define the model. We measure the time autocorrelation
function of the stress tensor, the shear viscosity, the structural
relaxation times and the mass diffusion coefficient.

We first determine the time autocorrelation functions of a structural
quantity which is well defined in all minimum configurations, i.e. the
off-diagonal microscopic stress tensor (\ref{tsfo1}). The correlation
function is 
\begin{equation}
C(t)=\frac 13[<\sigma ^{zx}(t)\ \sigma ^{zx}(0)>+(xy)+(yz)]\ .  \label{corre}
\end{equation}
The quantity ${\cal O}_{ab}$ in Eq. (\ref{me3}) is in this case 
\begin{equation}
{\cal O}_{ab}=\frac 13[<\sigma _a^{zx}\ \sigma _b^{zx}>+(xy)+(yz)]\ .
\end{equation}
We have measured the correlation functions for different temperatures, from 
$T=150$ K to $T=20$ K. In Fig. $7$ we report the normalized correlation
functions $C(t)/C(0)$ at different temperatures (open symbols) together with
the best stretched exponential fit (lines): 
\begin{equation}
C(t)=C(0) \; \exp \left[ -(t/\tau )^{\beta _k}\right] \ .  \label{ste}
\end{equation}
Contrary to the MD computations, which result in a two-step behavior for the
relaxation processes (one associated to fast local dynamics and the other to
structural slow dynamics, the so called $\alpha $ structural processes), the
model gives only one relaxation step, associated to the structural
processes, because the model can only describes the long time behavior. The
results we obtain with the present model are consistent with those of MD in
the allowed region ( i.e. above $T\sim 90$ K, in order to avoid
crystallization in the MD computation). In inset $(a)$ we show the
temperature dependence of the stretching parameter $\beta _k$. It emerges
that: the structural relaxation dynamics is well represented by a stretched
exponential decay, and that the stretching parameter $\beta _k$ is strongly
temperature dependent.

Both results are well supported by experimental \cite{exp_beta} and
numerical \cite{num_beta} observations. In our case the stretching
parameters $\beta _k$ decreases from a value of $\sim 1$ at high temperature
to $\beta _k\sim 0.35$ at low temperature, in agreement with experimental
findings \cite{exp_beta}.

From the behavior of the correlation function $C(t)$ we can obtain
information about structural relaxation time $\tau $. The values of $\tau $
(inset $(b)$) are determined from the stretched exponential fits. We obtain
an increase of many orders of magnitude in a small temperature range, as
usually found in many glass-forming liquids. However, we do not find the
dramatic increase of the Vogel-Tammann-Fulcher type expected for fragile
glass-former \cite{ANGELL}. It is possible that the observed Arrhenius
behavior emerges as a peculiar property of the model, which would mean that
the model is unable to capture the phenomenology of ``fragility''. It is,
however, possible that the Arrhenius law is a genuine property of
glass-forming liquids with Lennard-Jones interaction, as supported by a
comparison with a MD computation in the allowed temperature range. It would
be very interesting to compare the behavior obtained from the model to the
``true'' (in the sense of MD computation) behavior in the full temperature
range (this is possible only for those system that avoid crystallization,
like suitable binary mixtures).

From the time autocorrelation functions of the off-diagonal microscopic
stress tensor, we can determine the shear viscosity as \cite{BALUCANI} 
\begin{equation}
\eta =\frac 1{K_BTV}\int_0^\infty dt\ C(t)\ .
\end{equation}
Also in this case, like for the relaxation times, we find a strong increase
in a small temperature range, from $\eta \sim 10^{-2}$ P at $T=150$ K, to 
$\eta \sim 10^{11}$ P at $T=20$ K. Also in this case we found a good
agreement between the model and MD for $T>90$ K, giving further support to
the model.

The last quantity we measured in the model was the mass diffusion
coefficient. In order to find it we determine the mean square displacement 
\begin{equation}
{\cal O}(t)=\frac 1N|{\underline{r}}(t)-{\underline{r}}(0)|^2=\frac 1N
\sum_{i=1}^N|{\vec r}_i(t)-{\vec r}_i(0)|^2\ ,
\end{equation}
from which we obtain the diffusion coefficient: 
\begin{equation}
D=\lim_{t\rightarrow \infty }\ \frac{{\cal O}(t)}{6t}\ .
\end{equation}
To evaluate $D$ we use the quantity ${\cal O}_{ab}=d_{ab}^2/N$, where 
$d_{ab}$ is defined in (\ref{dista}).
We observe again a strong increase with an interesting behavior 
not simply linear in double logarithmic scale 
\cite{lettera}.

We conclude this section by analyzing the validity of the Stokes-Einstein
relation in the model. The Stokes-Einstein relation describes in a rigorous
way the diffusive motion of a macroscopic objects in a fluids and predicts
the following relation: 
\begin{equation}
D\propto \frac T\eta \ .  \label{se}
\end{equation}
The Stokes-Einstein relation also describes fairly well the diffusion at
atomic scale in liquids at high temperatures. By lowering temperature, as
observed in many experiments \cite{BSE}, one usually finds a breakdown of
the Eq. (\ref{se}). 
We found \cite{lettera} that at high $T$ the model satisfies 
asymptotically the Stokes-Einstein relation, 
but upon decreasing $T$ we observe a breakdown of
the relation and a fit over the lowest temperature data of the type 
\begin{equation}
D^{-1}\propto (\frac \eta T)^\xi \ ,
\end{equation}
gives the value 
\begin{equation}
\xi \simeq 0.28\ .
\end{equation}
This value is in fairly good agreement with experimental results found in
fragile glass-formers, like o-terphenyl \cite{FDSE}.

\hfill

\subsection{Off-equilibrium properties}

Although introduced to analyze the long time dynamics, the model allows an
easy computation of the short time, off-equilibrium dynamics. One of the
main properties of glass-formers is the very strong increase of
characteristic relaxation times when temperature is lowered. If these times
become comparable to observational times, the system is no more able to
explore the full accessible phase space and then to reach the thermal
equilibrium. The observed quantities are characterized by off-equilibrium
processes. In this regime the one-time quantities, such as energy or time
correlation functions with fixed initial time, can no longer describe the
physics of the system. The usual translational time invariance, valid in the
equilibrium regime, is no more satisfied. One of the most interesting
consequences of that is the fact that the fluctuation-dissipation relation 
does not longer hold\cite{BOUCH}. We concentrate here on this
property.

Let be $H$ the Hamiltonian of the system and ${\cal O}$ a generic observable
dependent on microscopic variables. We define the two time autocorrelation
function 
\begin{equation}
C(t,t_w)=<{\cal O}(t)\ {\cal O}(t_w)>\ ,  \label{cor1}
\end{equation}
where we suppose $t>t_w$ and $<...>$ now means a dynamical average over 
initial conditions. We also introduce the response function to a perturbation 
$\epsilon $, which is coupled to the observable ${\cal O}$ and gives rise to 
a pertutbed Hamiltonian: 
\begin{equation}
H^{\prime }=H+\epsilon (t)\ {\cal O}\ ,
\end{equation}
The response is defined as 
\begin{equation}
R(t,t_w)=\left. \frac{\delta <{\cal O}(t)>}{\delta \epsilon (t_w)}\right|
_{\epsilon =0}\ ,
\end{equation}
where again $t>t_w$. In the equilibrium regime the time translational
invariance implies the validity of fluctuation dissipation relation 
\cite{PARSTA} 
\begin{equation}
R_{eq}(\tau )=\beta \ \frac{\partial C_{eq}(\tau )}{\partial \tau }\ ,
\label{fde}
\end{equation}
where $\tau =t-t_w$. Introducing the integrate response function $\chi $ 
\begin{equation}
\chi (t,t_w)=\int_{t_w}^tdt^{\prime }\ R(t,t^{\prime })\ ,
\end{equation}
Eq. (\ref{fde}) takes the form 
\begin{equation}
\frac{d\chi (C)}{dC}=\beta \ .  \label{fdne3}
\end{equation}
In the off-equilibrium regime the fluctuation-dissipation ratio (\ref{fdne3})
is no more valid. It is possible however to generalize the ratio
introducing a violation factor $X(t,t_w)$. The analytical study of some
generalized mean field spin glass models \cite{CUGLI} shows that the
function $X(t,t_w)$ depends on time only through the correlation function 
$C$: $X(t,t_w)=X[C(t,t_w)]$. Using this property we can write a generalized
fluctuation-dissipation ratio in the off-equilibrium regime 
\begin{equation}
\frac{d\chi (C)}{dC}=\beta \ X(C)\ .  \label{fdne4}
\end{equation}
For short times $\tau \ll t_w$ we have $X(C)=1$ and the system satisfies an
equilibrium-like relation, even if it is confined in a small phase space 
region. For times $\tau \sim t_w$ the exploration of the phase space is an
off-equilibrium process and this implies the violation of the equilibrium
fluctuation-dissipation ratio. In this case we have $X(C)<1$. The very
interesting relationship between off-equilibrium and equilibrium properties
of some generalized spin glass model, suggests that in the case of one step
replica symmetry breaking the $X(C)$ function depends only on temperature 
\begin{equation}
X(C)=m(T)\ ,
\end{equation}
and $m(T)$ is linear in $T$ at low temperature. It has been recently
suggested that structural glasses present a striking similarity with the
generalized spin glass model with one step replica symmetry breaking \cite
{equiv} (for a recent interesting review see Coluzzi \cite{COLUZZI}). We
then expect that also for structural glasses the violation parameter would
show a linear temperature dependence in the violation region $X<1$. Evidence
of this behavior was found in recent numerical study of binary mixtures \cite
{PABA}; we analyze it in our model.

Let be ${\cal O}({\underline{r}}(t))$ a generic observable that depends on
collective coordinates at time $t$. The average value of ${\cal O}$ in the
off-equilibrium regime in the model is 
\begin{equation}
<{\cal O}(t)>=\frac 1{M^{\prime }}\sum_{a,b}{\cal O}_a\ p_a(t;b,0)\ ,
\label{medne}
\end{equation}
where $a$ and $b$ label the minima and the sum over $b$ is now limited to a
certain subset of minima. We chose the $M^{\prime
}$ highest energy states. Expression (\ref{medne}) differs from the
equilibrium one, as the initial states are weighted with a constant term
(corresponding to an infinite temperature) rather than with the
Gibbs-Boltzmann equilibrium weight. In this way we describe an instantaneous
quench at time $t=0$ from $T=\infty $ to a finite temperature $T$ (the $T$
dependence is as usual in the probability $p_a$). The sum restricted to the 
$M^{\prime }$ initial states with highest energies ($M^{\prime }<M$ where $M$
is the total number of minima; in our case $M=400$ and $M^{\prime }=20$)
allows a better description of the off-equilibrium regime. We calculate the
time correlation functions in the model as 
\begin{equation}
<{\cal O}(t)\ {\cal O}(t_w)>=\frac 1{M^{\prime }}\sum_{a,b,c}{\cal O}_a\ 
{\cal O}_b\ p_b(t_w;c,0)\ p_a(t;b,t_w)\ ,  \label{corne}
\end{equation}
where the sum over $b$ is still made over the $M^{\prime }$ highest minima.
The quantity we determine is the time autocorrelation function of the
off-diagonal microscopic stress tensor $\sigma ^{zx}$: 
\begin{equation}
C(t,t_w)=<\sigma ^{zx}(t)\ \sigma ^{zx}(t_w)>-<\sigma ^{zx}(t)>\ <\sigma
^{zx}(t_w)>\ .
\end{equation}
The response function is determined by the perturbed Hamiltonian 
\begin{equation}
H^{\prime }=H+\epsilon (t)\ \sigma ^{zx}\ ,
\end{equation}
where the external field $\epsilon $ is 
\begin{equation}
\epsilon (t)=\left\{ 
\begin{array}{ll}
\displaystyle 0\displaystyle \ \ \ \ \ \ \ \  & \mbox{for \ \ $t<t_w$}\ , \\ 
&  \\ 
\displaystyle \epsilon \displaystyle \ \ \ \ \ \ \ \  & 
\mbox{for \ \ $t\geq
t_w$}\ .
\end{array}
\right.   \label{pert}
\end{equation}
The perturbation induces a change in the energies of the minima: 
\begin{equation}
\Phi _a^{^{\prime }}=\Phi _a+\epsilon (t)\ \sigma _a^{zx}\ .
\end{equation}
The response function is 
\begin{equation}
R(t,t_w)=\left. \frac{\delta <\sigma ^{zx}(t)>_\epsilon }{\delta \epsilon
(t_w)}\right| _{\epsilon =0}\ .
\end{equation}
The $<...>_\epsilon $ is evaluated in the presence of the perturbation 
$\epsilon $: 
\begin{equation}
<\sigma ^{zx}(t)>_\epsilon =\frac 1{M^{\prime }}\sum_{a,b,c}\sigma _a^{zx}\
p_b(t_w;c,0)\ p_a^\epsilon (t;b,t_w)\ ,  \label{medper}
\end{equation}
where $p^\epsilon $ is the solution of the master equation with the
perturbing term. For small perturbation we obtain: 
\begin{equation}
\chi (t,t_w)=\frac{<\sigma ^{zx}(t))>_\epsilon -<\sigma ^{zx}(t))>_{\epsilon
=0}}\epsilon \ .
\end{equation}

We have determined the correlation functions $C(t,t_w)$ and the response 
$\chi (t,t_w)$ as functions of $t$ for different times $t_w$; the
temperatures we analyze are in the range $T=100\div 20$ K. In determining
the response functions we have used a value of $\epsilon $ small enough 
($\epsilon =0.1$) that the regime is linear, as verified by trying different 
$\epsilon $ values. In Fig. $8$ we report the behavior of $\chi $
versus $\beta C$ at temperatures $T=90$ K and $T=45$ K, respectively.
While at $T=90$ K the relation between $\chi $ and $\beta C$ is to a good
approximation linear with slope $1$ on the whole range (full line), at lower
temperature it is evident that after a first linear behavior with slope $1$
(full line)  an approximately linear behavior with slope $<1$ takes on at
longer times (dashed line), as theoretically and numerically expected.
Moreover the slope of the second region decreases by lowering temperature:
in Fig. $9$ we show the slope $m$ of the violation region versus $T$. At
high temperature the value of $m$ is nearly $1$, while below a temperature
of about $60\div 70$ K $m$ decreases linearly, as we expect in the
hypothesis of one step replica symmetry breaking. Fig. $9$ is limited to  
$T>40$ K, as for lower temperatures the $m$ values saturate to a limiting value
and it is no more possible to extract correct information. 
This effect is probably due to the finite size of the system,
because the sampling of the initial off-equilibrium states is not exhaustive
($M^{\prime }=20$). In the equilibrium analysis the finite size effects do
not show up in the temperature range explored, as the sampling of the
initial states is  complete ($M^{\prime }=M$).

In conclusion from the analysis of the off-equilibrium properties of the
model it emerges that the deviation from the usual
fluctuation-dissipation relation, valid in the equilibrium regime, is in
agreement with theoretical predictions and numerical findings in simple
glass-formers.


\section{Conclusions}

The very rich phenomenology of the cooling process of glass-forming liquids,
of the glass transition and of the glassy systems in general, has received
in the last few years many important theoretical, numerical and experimental
contributions. The present work is concerned with  the numerical
investigation of a simple model glass, a Lennard-Jones system of interacting
particles. The main aim of the work was to determine the emergent properties
of the system at the level of potential energy landscape. After a detailed
analysis of the topological properties of the potential energy surface, we
introduced a model which reproduces the long time dynamic behavior of the
system. While in the usual MD investigations of relaxation the computational
times are proportional to physical times (with computational times of the
order $10^5$ s one obtains physical times of the order $10^{-9}$ s 
for system of size $N\sim 10^3$), our model allows the study at very 
long physical times in short computational times.

We studied both equilibrium and off-equilibrium properties. The main
equilibrium results we obtained are ({\it i}) the stretching of the
relaxation dynamics, ({\it ii}) the temperature dependence of the stretching
parameter, and ({\it iii}) the breakdown of the Stokes-Einstein relation. If
they are genuine properties of the glassy system analyzed, they represent
intriguing and interesting results that open fascinating questions about the
glassy and supercooled liquid behavior.

Although introduced to investigate the long time dynamics, the model is also
able to describe in a simple and direct way the off equilibrium dynamics.
The emergent violation of the fluctuation-dissipation relation (that holds
at equilibrium) is a very interesting feature and  supports many conjectures
about the analogy between structural glasses and some spin glass model 
\cite{equiv}. 
Moreover, the appearance of a
critical temperature below which the violation takes place, seems to
indicate the existence of a transition; its relation with the equilibrium
properties is an open and exciting question.

In conclusion, the analyzed features of the potential energy landscape and
the emergent properties of the model both at and off equilibrium, seem to
provide a good description of the glassy systems.  The method is very
powerful for the investigation of the glassy properties by avoiding some of
the main problems usually encountered in numerical studies, like the very
long computational times in the low temperature regime or the presence of
crystal states. We hope the analysis we performed may constitute a promising
route in the investigation of glassy systems.

\hfill

We acknowledge B.~Coluzzi, G.~Monaco, F.~Sciortino and P.~Verrocchio for 
useful discussions, and D.~Leporini who kept our attention on the fractional 
SE relation issue.


\newpage

\vskip 1cm
\noindent
\narrowtext
\begin{table*}
\caption{Correlation coefficients $r$ for the measured quantities 
$x$ and $y$.}
\begin{tabular}{||c|c|c|c||}
$N$ & $x$ & $y$ & $r$ \\
\hline
\hline
$29$ &  $\Phi_a$  &  $c_a$  &  $0.13$  \\
\hline
$29$ &  $\Phi_a$  &  $\gamma_a$   & $0.46$  \\
\hline
$29$ &  $\vert \Phi_a - \Phi_b \vert$  &  $d_{ab}$  &  $0.12$ \\
\hline
$29$ & $\ln \vert \Phi_a - \Phi_b \vert$ & $\ln d_{ab}$   & $0.17$ \\
\hline
$17$ &  $\vert \Phi_a - \Phi_b \vert$  & $d_{ab}$  &  $0.18$ \\
\hline
$17$ & $\ln \vert \Phi_a - \Phi_b \vert$  &  $\ln d_{ab}$  & $-0.11$ \\
\hline
$17$ &  $\Phi_{sad.}$  &  $c_{sad.}$  &  $0.23$  \\
\hline
$17$ &  $\Phi_{sad.}$  &  $\gamma_{sad.}$   & $0.49$  \\
\hline
$17$ &  $c_{sad.}$  &  $\tilde{\omega}_{sad.}$  & $-0.12$  \\
\hline
$17$ &  $\gamma_{sad.}$  & $\ln \tilde{\omega}_{sad.}$  & $-0.31$  \\
\hline
$17$ &  $d_{ab}$  &  $\Delta\Phi_{bar.}$  &  $0.47$  \\
\hline
$17$ & $\ln d_{ab}$  &  $\ln \Delta\Phi_{bar.}$   & $0.67$  \\
\end{tabular}
\end{table*}

{\footnotesize{
\begin{center}
{\bf FIGURE CAPTIONS}
\end{center}

\begin{description}

\item  {Fig. 1 - 
Distribution of the potential energy (in K per particle)
of the minima for the system $N=29$.
In the inset the distribution of the curvature $\gamma_{min.}$.
}

\item  {Fig. 2 - 
An example of potential energy profile along the 
least action path ($\bullet$) between two minima,
compared with the straight path ($\circ$).
}

\item  {Fig. 3 - 
Energy distribution of the barriers along the least action paths 
among the minima.
The inset shows the distribution of the curvature $\gamma_{sad.}$
of the saddle points.}

\item  {Fig. 4 - 
Correlation between energies and curvatures of the minima.
The highest curvature values correspond to low energy crystal minima. 
}

\item  {Fig. 5 - 
Correlation between energies of the barriers and distances among minima,
in a $\log$ scale.
The line is the best linear fit in $\log$ scale, 
correspondent to a power law fit (the slop is $\alpha = 3.7$).
}

\item  {Fig. 6 - 
Potential energy versus temperature as determined from MD
and from the model.
The MD ($\circ$) are obtained heating the glass and 
($\bullet$) cooling the liquid.
Dotted line refers to the model using all the minima, while full line
using only the glassy minima.
}

\item  {Fig. 7 - 
Normalized autocorrelation functions of the off-diagonal microscopic stress 
tensor versus time at different temperatures, obtained from the model. 
In the inset $(a)$  the $T$ dependence of the stretching parameter $\beta_K$
and in the inset $(b)$ the relaxation time versus $1/T$, both obtained from
the 
stretched exponential fit of the autocorrelation functions.
}

\item  {Fig. 8 - 
The integrate response $\chi$ versus $\beta C$ at temperatures 
$T=90$ K ($\circ$) and $T=45$ K ($\bullet$).
The full line is the fluctuation-dissipation ratio, 
the dotted line is the best fit of the last points of $T=45$ K. 
}

\item  {Fig. 9 - 
The slop $m$ in the region of the violation of the fluctuation-dissipation 
ratio versus temperature.
The straight line fits the data in the violation region. 
}
\end{description}

}
}
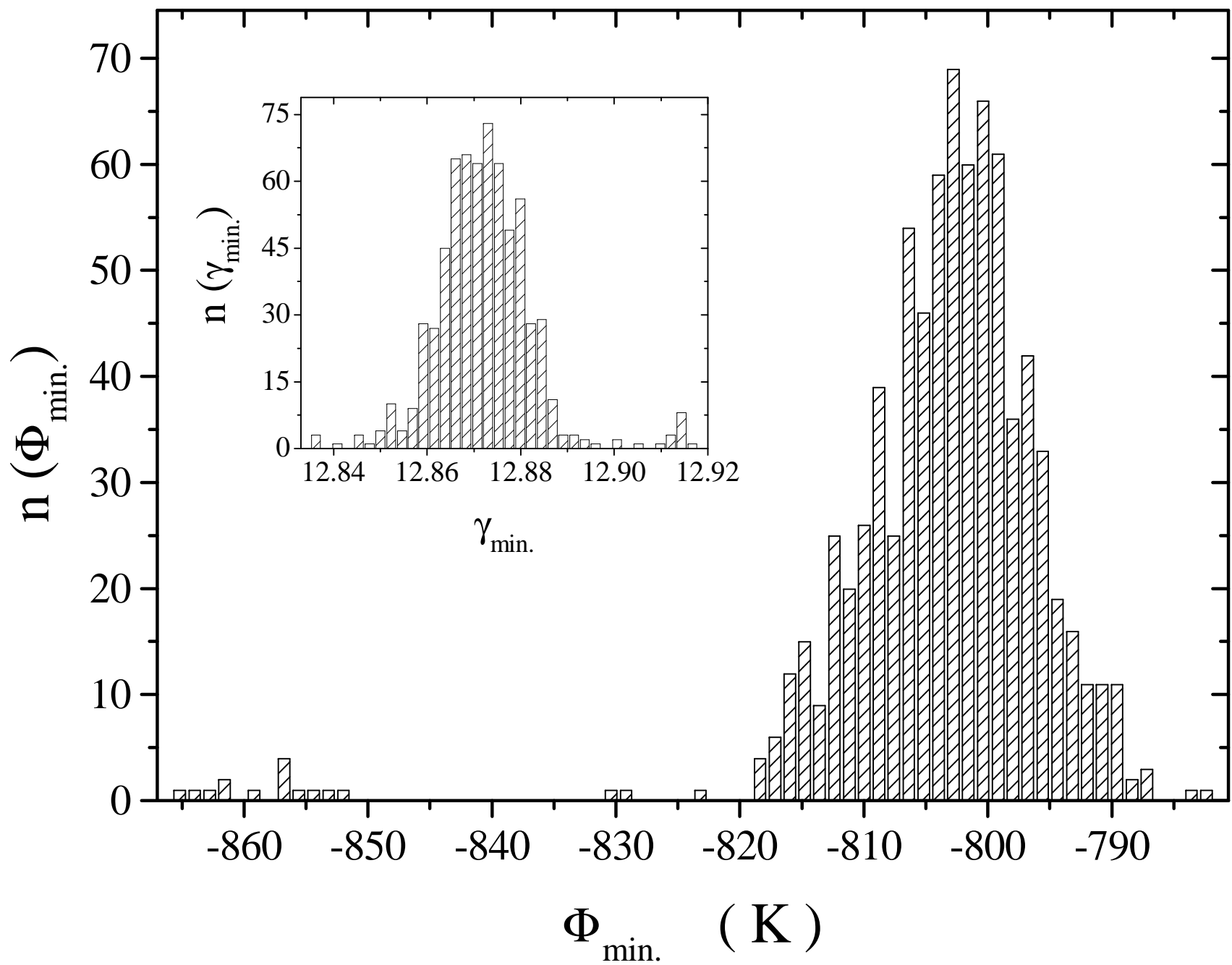
\end{document}